\documentclass[a4paper,12pt]{article}
\usepackage{graphicx}
\usepackage{amsthm}
\usepackage{amsfonts}
\usepackage{amsmath,latexsym}
\usepackage{amssymb}
\usepackage{hyperref}
\usepackage{tabularx}
\newcolumntype{M}{>{$}c<{$}}
\usepackage[matrix,arrow,curve]{xy}

\numberwithin{equation}{section} \numberwithin{figure}{section}
\numberwithin{table}{section}

\def\papertitlepage{\baselineskip 3.5ex\thispagestyle{empty}}
\def\Title#1{\baselineskip 1cm \vspace{1.5cm}%
  \begin{center}{\Large\bf #1}\end{center}\vspace{0.5cm}}
\def\Authors#1{\begin{center}\renewcommand{\thefootnote}{\fnsymbol{footnote}}{\it #1}\end{center}}
\def\Abstract{\vspace{1.0cm}%
  \begin{center}{\large\bf Abstract}\end{center}}

\renewenvironment{thebibliography}{\pagebreak[3]\par\vspace{0.6em}
\begin{flushleft}{\large \bf References}\end{flushleft}
\vspace{-1.0em}

\begin{enumerate}\if@twocolumn\baselineskip=0.6em\itemsep -0.2em
\else\itemsep -0.2em\fi\labelsep 0.1em}{\end{enumerate} }

\begin{document}
{\papertitlepage \vspace*{0cm} {\hfill
\begin{minipage}{4.2cm}
CCNH-UFABC 2021\par\noindent April, 2021
\end{minipage}}
\Title{$KBc$ algebra and the gauge invariant overlap in open
string field theory}
\Authors{{\sc E.~Aldo~Arroyo${}$\footnote{\tt
aldo.arroyo@ufabc.edu.br}}
\\
Centro de Ci\^{e}ncias Naturais e Humanas, Universidade Federal do ABC \\[-2ex]
Santo Andr\'{e}, 09210-170 S\~{a}o Paulo, SP, Brazil ${}$ }
} 

\vskip-\baselineskip
{\baselineskip .5cm \Abstract We study in detail the evaluation of
the gauge invariant overlap for analytic solutions constructed out
of elements in the $KBc$ algebra in open string field theory. We
compute this gauge invariant observable using analytical and
numerical techniques based on the sliver frame $\mathcal{L}_0$ and
traditional Virasoro $L_0$ level expansions of the solutions.
 }
\newpage
\setcounter{footnote}{0}
\tableofcontents

\section{Introduction}

It is well-known that the analytic solutions for tachyon
condensation
\cite{Schnabl:2005gv,Erler:2009uj,Schnabl:2010tb,Okawa:2006vm,Arroyo:2010sy,Jokel:2017vlt,Arroyo:2017mpd}
in open bosonic string field theory \cite{Witten:1985cc} can be
formally given in terms of elements in the $KBc$ algebra
\cite{Erler:2006hw,Erler:2006ww}. Once a solution $\Psi$ is given,
the next step is to evaluate relevant physical gauge invariant
quantities, like the energy and the gauge invariant overlap
$\langle I | \mathcal{V}(i) | \Psi \rangle$ discovered in
\cite{Hashimoto:2001sm, Gaiotto:2001ji,Kawano:2008ry}. As argued
by Ellwood \cite{Ellwood:2008jh}, the gauge invariant overlap
represents the shift in the closed string tadpole of the solution
relative to the perturbative vacuum. Moreover, using an
appropriate zero momentum vertex operator $\mathcal{V}$, defined
in \cite{Baba:2012cs}, it has been shown that the value of the
energy can be obtained from the gauge invariant overlap.

The analytic computation of the gauge invariant overlap for
Schnabl's tachyon vacuum solution has been performed in
\cite{Kawano:2008ry}. Although the evaluation of this gauge
invariant appears to be simpler than the energy, the computation
presented in \cite{Kawano:2008ry} was a bit cumbersome, and the
reason for this subtlety was that the authors used a
representation of the solution as given in Schnabl's original work
\cite{Schnabl:2005gv}. As we will see, the computation of the
gauge invariant overlap can be enormously simplified if we express
Schnabl's solution in terms of elements in the $KBc$ algebra.

Concerning the numerical analysis of the gauge invariant overlap
for analytic solutions within the $KBc$ algebra, in reference
\cite{Kawano:2008ry}, using the traditional Virasoro $L_0$ level
truncation scheme the authors have evaluated the gauge invariant
overlap for the case of Schnabl's original solution. Regarding the
case of the so-called Erler-Schnabl's solution, although the
analytical computation of the gauge invariant overlap for this
solution has been performed in \cite{Erler:2009uj}, up to now,
using the Virasoro $L_0$ level truncation scheme, the analysis of
the gauge invariant overlap for this type of solution was not
performed. Moreover, the analysis of the gauge invariant overlap
by means of the curly $\mathcal{L}_0$ level truncation scheme has
not been carried out neither for Schnabl nor for Erler-Schnabl's
solution. And in the case of the new real tachyon vacuum solution
discovered in reference \cite{Jokel:2017vlt} (called as Jokel's
solution \cite{Arroyo:2017mpd}), neither the numerical nor the
analytical computation was presented for the gauge invariant
overlap.

Motivated by the above results and open issues, in this work,
using analytical and numerical techniques based on the curly
$\mathcal{L}_0$ and the traditional Virasoro $L_0$ level
truncation schemes, we show a detailed and pedagogical way of
computing the gauge invariant overlap for solutions constructed
out of elements in the $KBc$ algebra. As explicit examples of our
generic results, we present the analytical and numerical
computation of the gauge invariant overlap for Schnabl's,
Erler-Schnabl's and Jokel's solutions.

By expanding the solutions $\Psi$ in the basis of curly
$\mathcal{L}_0$ eigenstates, we surprisingly discover that the
result for the gauge invariant overlap $\langle I | \mathcal{V}(i)
| \Psi \rangle$ turns out to be a finite series. This result is in
contrast to the case of the energy, where the series has an
infinite number of terms and diverges, though this divergent
series can be resummed numerically by means of Pad\'{e}
approximants to give a good approximation to the expected value of
the D-brane tension
\cite{Schnabl:2005gv,Erler:2009uj,Arroyo:2009ec,AldoArroyo:2011gx}.

Regarding the numerical result of the gauge invariant overlap for
Erler-Schnabl's and Jokel's solution obtained by means of Virasoro
$L_0$ level truncation computations, we would like to mention that
the main reason for performing this numerical computation is to
see whether or not higher level contributions yield to
increasingly convergent results which approach the expected
answer. We will show that the series that represents the gauge
invariant overlap for these solutions turn out to be a divergent
ones, therefore we will be required to use Pad\'{e} approximants.

This paper is organized as follows. In section 2, we introduce the
sliver frame and discuss some conventions and definitions that
will be used in the rest of the paper. In section 3, we review the
$KBc$ algebra. In sections 4, 5 and 6, we evaluate analytically
and numerically the gauge invariant overlap for solutions
expressed in terms of elements in the $KBc$ algebra. In section 7,
a summary and further directions of exploration are given.

\section{The sliver frame: conventions and definitions}
Originally, the sliver frame has been defined as the $\tilde z$ coordinate obtained by the map \cite{Schnabl:2005gv}
\begin{eqnarray}
\label{map1} \tilde z  = \arctan z,
\end{eqnarray}
where $z$ is a point on the upper half-plane (UHP). It is known
that the gluing prescription entering into the definition of the
star product simplifies if one uses the $\tilde z$ coordinate.
Under the map (\ref{map1}), the UHP looks as a semi-infinite
cylinder of circumference $\pi$ denoted by $C_{\pi}$.

There is another convention for the definition of the sliver frame
which uses the following map
\begin{eqnarray}
\label{map2} \tilde z = \frac{2}{\pi} \arctan z.
\end{eqnarray}
This map has been used in reference \cite{Erler:2009uj}, and in
this case, the UHP looks as a semi-infinite cylinder of
circumference $2$ denoted by $C_{2}$.

Since the expressions written in terms of elements in the $KBc$
algebra which are used in the construction of analytic solutions
look different depending on the convention adopted for the $\tilde
z$ coordinate, it is always useful to mention, from the beginning,
what of those conventions will be chosen, i.e., the one given by
(\ref{map1}) or (\ref{map2}).

In the literature some authors use the convention (\ref{map1}) and
others (\ref{map2}), in this work we are going to use a rather
generic definition which takes into account these two conventions.
Let us define the $\tilde z$ coordinate by the map
\begin{eqnarray}
\label{map3} \tilde z = \frac{l}{\pi} \arctan z,
\end{eqnarray}
so that the UHP looks as a semi-infinite cylinder of circumference
$l$ denoted by $C_{l}$. Note that the case $l=\pi$ corresponds to
the convention (\ref{map1}) while the case $l=2$ corresponds to
(\ref{map2}).

Let us define the operators $\hat{\mathcal{L}}$,
$\hat{\mathcal{B}}$ and $\tilde c_p$ which are very useful in the
construction of elements in the $KBc$ algebra. These operators are
related to the worldsheet energy-momentum tensor $T$, the $b$ and
$c$ ghosts fields respectively. Using the map (\ref{map3}), we can
write the explicit definition of the operators
$\hat{\mathcal{L}}$, $\hat{\mathcal{B}}$ and $\tilde c_p$
\begin{align}
\label{Lhat01} \hat{\mathcal{L}} \equiv \mathcal{L}_{0} + \mathcal{L}^{\dag}_0 &= \oint \frac{d z}{2 \pi i} (1+z^{2}) (\arctan z+\text{arccot} z) \,
T(z) \, , \\
\label{Bhat01} \hat{\mathcal{B}} \equiv \mathcal{B}_{0} + \mathcal{B}^{\dag}_0 &= \oint \frac{d z}{2 \pi i} (1+z^{2}) (\arctan z+\text{arccot} z) \, b(z)
\, , \\
\label{c01} \tilde c_p &= \Big(\frac{l}{\pi}\Big)^{p} \oint \frac{d z}{2 \pi i} \frac{1}{(1+z^2)^2} (\arctan z)^{p-2} c(z).
\end{align}
In general, if we have a primary field $\phi$ with conformal weight $h$, using the map (\ref{map3}), we obtain
\begin{eqnarray}
\label{phi01} \tilde \phi_p \equiv \oint \frac{d \tilde z}{2 \pi i} \tilde z^{p+h-1} \tilde \phi (\tilde z)  = \Big(\frac{l}{\pi}\Big)^{p} \oint \frac{d
z}{2 \pi i} \frac{1}{(1+z^2)^{1-h}} (\arctan z)^{p+h-1} \phi(z).
\end{eqnarray}
Using this equation (\ref{phi01}), let us define the operators $\mathcal{L}_{-1}$ and $\mathcal{B}_{-1}$ which are useful in the computation of the star
product of string fields involving the operators $\hat{\mathcal{L}}$ and $\hat{\mathcal{B}}$
\begin{align}
\label{Lm101} \mathcal{L}_{-1}  \equiv \oint \frac{d \tilde z}{2 \pi i} \tilde T(\tilde z) &= \frac{\pi}{l} \oint \frac{d z}{2 \pi i} (1+z^{2})  \,
T(z) =  \frac{\pi}{l} \big[L_{-1} + L_{1} \big] \, , \\
\label{Bm101} \mathcal{B}_{-1} \equiv \oint \frac{d \tilde z}{2 \pi i} \tilde b(\tilde z) &= \frac{\pi}{l} \oint \frac{d z}{2 \pi i} (1+z^{2})  \, b(z)\;
=\frac{\pi}{l} \big[b_{-1} + b_{1} \big] \, .
\end{align}

To compute the star product of string fields involving the operators $\hat{\mathcal{L}}$, $\hat{\mathcal{B}}$ and $\tilde c_p$, we will need to know the
following commutator and anti-commutator relations
\begin{align}
\label{com01} [\mathcal{L}_{-1},\hat{\mathcal{L}}] = [\mathcal{L}_{-1},\hat{\mathcal{B}}] = 0 \; , \;\; [\mathcal{L}_{-1},\tilde c_p] = (2-p) \tilde
c_{p-1} \, . \\
\label{com02} [\hat{\mathcal{B}},\hat{\mathcal{L}}] = [\mathcal{B}_{-1},\hat{\mathcal{L}}] = \{\mathcal{B}_{-1},\hat{\mathcal{B}}\} = 0  \; , \;
\{\mathcal{B}_{-1},\tilde c_p\} = \delta_{p-1,0} \, .
\end{align}

To represent the elements in the $KBc$ algebra, we will require to
know the operator $U^{\dag}_r U_r$. This operator can be written
in terms of the operator $\hat{\mathcal{L}}$
\begin{eqnarray}
\label{Ur01} U^{\dag}_r U_r = \exp \Big[ \frac{2-r}{2} \hat{\mathcal{L}}\Big].
\end{eqnarray}

\section{Star products and the $KBc$ algebra}
Before defining the basic elements belonging to the $KBc$ algebra,
we are going to write the star product of string fields containing
the operators $\hat{\mathcal{L}}$ and $\hat{\mathcal{B}}$. Given
two string fields $\phi_1$ and $\phi_2$, we can show that
\begin{align}
\label{pro1} (\hat{\mathcal{B}}\phi_1)*\phi_2 &= \hat{\mathcal{B}}(\phi_1*\phi_2) +
(-1)^{\text{gh}(\phi_1)} \frac{l}{2} \phi_1 * \mathcal{B}_{-1} \phi_2 \; , \\
\label{pro2} \phi_1 * (\hat{\mathcal{B}}\phi_2)&=
(-1)^{\text{gh}(\phi_1)}\hat{\mathcal{B}}(\phi_1*\phi_2)-(-1)^{\text{gh}(\phi_1)}\frac{l}{2}(\mathcal{B}_{-1}\phi_1)*\phi_2
\; ,  \\
\label{pro3} (\hat{\mathcal{B}}\phi_1)*(\hat{\mathcal{B}}\phi_2)&=-(-1)^{\text{gh}(\phi_1)}
\frac{l}{2}\hat{\mathcal{B}}\mathcal{B}_{-1}(\phi_1*\phi_2)+(\frac{l}{2})^2
(\mathcal{B}_{-1} \phi_1)*(\mathcal{B}_{-1} \phi_2)\; , \\
\label{pro4} (\hat{\mathcal{L}}^{n}\phi_1)*\phi_2 &=
\sum_{n'=0}^{n}  {n \choose n'} (\frac{l}{2})^{n'} \hat{\mathcal{L}}^{n-n'}(\phi_1*\mathcal{L}_{-1}^{n'}\phi_2)\; , \\
\label{pro5} \phi_1*(\hat{\mathcal{L}}^{n}\phi_2) &= \sum_{n'=0}^{n}  {n \choose n'} (\frac{l}{2})^{n'} (-1)^{n'}
\hat{\mathcal{L}}^{n-n'}((\mathcal{L}_{-1}^{n'}\phi_1)*\phi_2)\; , \\
\label{pro6} (\hat{\mathcal{L}}^{m}\phi_1)*(\hat{\mathcal{L}}^{n}\phi_2) &= \sum_{m'=0}^{m}\sum_{n'=0}^{n} {m \choose m'} {n \choose n'}
(\frac{l}{2})^{m'+n'} (-1)^{n'} \hat{\mathcal{L}}^{m+n-m'-n'}((\mathcal{L}_{-1}^{n'}\phi_1)*(\mathcal{L}_{-1}^{m'}\phi_2))\; ,
\end{align}
where $\text{gn}(\phi)$ takes into account the Grassmannality of
the string field $\phi$. If we set $l=\pi$, the above results
match the results given in reference \cite{Schnabl:2005gv}.

The action of the BRST, $\mathcal{L}_{-1}$, and $\mathcal{B}_{-1}$ operators on the star product of two string fields are given by
\begin{eqnarray}
\label{Q1} Q (\phi_1*\phi_2) &=& (Q \phi_1)*\phi_2 + (-1)^{\text{gn}(\phi_1)}\phi_1 *(Q
\phi_2), \\
\label{L1} \mathcal{L}_{-1} (\phi_1*\phi_2) &=& (\mathcal{L}_{-1} \phi_1)*\phi_2 + \phi_1 *(\mathcal{L}_{-1} \phi_2),
\\
\label{B1} \mathcal{B}_{-1} (\phi_1*\phi_2) &=& (\mathcal{B}_{-1} \phi_1)*\phi_2 + (-1)^{\text{gn}(\phi_1)}\phi_1 *(\mathcal{B}_{-1} \phi_2) \, .
\end{eqnarray}

Given a operator $\tilde \phi (\tilde z)$ defined in the $\tilde z$ coordinate, let us write the wedge state with insertion as follows
\begin{eqnarray}
\label{wed1} U_r^\dag U_r \tilde \phi (\tilde z)|0\rangle ,
\end{eqnarray}
where $U_r=(2/r)^{\mathcal{L}_0}$ is the scaling operator in the $\tilde z$ coordinate. The star product of two states $U_r^\dag U_r \tilde \phi (\tilde
x)|0\rangle$ and $U_s^\dag U_s \tilde \psi (\tilde y)|0\rangle$ can be derived using the usual gluing prescription
\begin{align}
\label{proweds1} U_r^\dag U_r \tilde \phi(\tilde x)|0\rangle
* U_s^\dag U_s \tilde \psi (\tilde y)|0\rangle= U_{r+s-1}^\dag
U_{r+s-1} \tilde \phi(\tilde x + \frac{l}{4}(s-1)) \tilde \psi(\tilde y-\frac{l}{4}(r-1))|0\rangle \, ,
\end{align}
where by $\tilde \phi (\tilde z)$ we have denoted the local operator $\phi (z)$ expressed in the sliver frame. For instance, in the case of a primary field
with conformal weight $h$, $\tilde \phi (\tilde z)$ is given by
\begin{eqnarray}
\label{conf1} \tilde{\phi}(\tilde{z}) = \big(\frac{dz}{d\tilde{z}}\big)^h \phi (z) = \big(\frac{\pi}{l}\big)^h \cos^{-2h}\big( \frac{\pi \tilde{z}}{l}
\big) \phi \Big( \tan \big( \frac{\pi \tilde{z}}{l} \big)  \Big).
\end{eqnarray}

The elements in the $KBc$ algebra are constructed out of the basic
string fields $K$, $B$, and $c$. These fields can be represented
in terms of operators acting on the identity string field $
|I\rangle=U_{1}^\dag U_{1} |0\rangle$
\begin{eqnarray}
\label{KK} K &\equiv& \frac{1}{l} \hat{\mathcal{L}} U_{1}^\dag U_{1} |0\rangle,
\\
\label{BB} B &\equiv& \frac{1}{l} \hat{\mathcal{B}} U_{1}^\dag U_{1} |0\rangle,
\\
\label{cc} c &\equiv&   U_{1}^\dag U_{1} \tilde c (0)|0\rangle.
\end{eqnarray}

Let us derive the algebra associated to the set of operators defined by equations (\ref{KK})-(\ref{cc}). As a pedagogical illustration, we explicitly
compute $\{B,c\}$
\begin{eqnarray}
\label{Bc1} \{B,c\} \equiv Bc + cB = \frac{1}{l} \hat{\mathcal{B}} U_{1}^\dag U_{1} |0\rangle * U_{1}^\dag U_{1} \tilde c (0) |0\rangle + \frac{1}{l}
U_{1}^\dag U_{1}\tilde c (0) |0\rangle * \hat{\mathcal{B}} U_{1}^\dag U_{1} |0\rangle,
\end{eqnarray}
using equations (\ref{pro1}), (\ref{pro2}) and the anti-commutator (\ref{com02}), we obtain
\begin{eqnarray}
\label{Bc2} \{B,c\} = U_{1}^\dag U_{1} |0\rangle = |I\rangle ,
\end{eqnarray}
therefore, we have that $\{B,c\}=1$.

Following the same steps, using equations (\ref{pro1})-(\ref{pro6}), the commutator and anti-commutator relations (\ref{com01}), (\ref{com02}), we can show
that
\begin{align}
\label{KBcal} [K,B]=0, \;\;\; \{B,c\} = 1, \;\;\; \partial c = [K,c],  \;\;\; B^2 = 0, \;\;\; c^2 = 0,
\end{align}
where the expression $\partial c$ has been defined as $\partial c \equiv U_{1}^\dag U_{1}
\partial \tilde c (0) |0\rangle $.

The action of the BRST operator $Q$ on the basic string fields $K$, $B$, and $c$ is given by
\begin{align}
\label{QKBc} Q K =0, \;\;\; Q B =K, \;\;\; Q c = cKc.
\end{align}

Employing the elements in the $KBc$ algebra, we can construct a
rather generic solution
\begin{align}
\label{Sol01} \Psi = Fc \frac{KB}{1-F^2}cF,
\end{align}
which formally satisfies the string field equation of motion $Q\Psi+\Psi \Psi =0$. For this solution to be a well defined string field, the function $F(K)$
must satisfy some holomorphicity conditions stated in reference \cite{Schnabl:2010tb}. From now, we will assume that $\Psi$ belongs to the set of well
defined string fields.

Let us list some solutions of the form (\ref{Sol01}). As a first
example, consider the analytic solution for the tachyon vacuum
\cite{Schnabl:2005gv}, where $F(K)=e^{-lK/4}$, actually Schnabl's
original solution corresponds to the case where $l=\pi$. Recall
that in this work, we are considering the map $\tilde z =
\frac{l}{\pi} \arctan z $, and therefore the Schnabl's solution
looks like
\begin{align}
\label{Sol02} \Psi_{\text{Sch}} = e^{-lK/4}c \frac{KB}{1-e^{-lK/2}}ce^{-lK/4}.
\end{align}
There is a subtlety with this solution, as shown in references \cite{Schnabl:2005gv,Schnabl:2010tb} when one performs the expansion of $K/(1-e^{-lK/2})$ as
the sum $\sum_n K e^{-lKn/2}$, the truncation of this sum produces a remnant which still contributes to certain observable \cite{Okawa:2006vm}. This is the
origin of the phantom term $\Psi_N$. Taking into account the phantom term, the solution (\ref{Sol02}) can be written as
\begin{align}
\label{Sol03} \Psi_{\text{Sch}} = \frac{2}{l} \lim_{N\rightarrow \infty} \Big[ \psi_N -\sum_{n=0}^{N-1} \frac{d \psi_n}{dn} \Big],
\end{align}
where
\begin{eqnarray}
\label{psin1} \psi_n = e^{-lK/4} cB e^{-lK n/2 } c e^{-lK/4}.
\end{eqnarray}

As a second example, let us consider the solution discovered by Erler and Schnabl, namely, the so-called simple tachyon vacuum solution \cite{Erler:2009uj}
\begin{align}
\label{Sol04} \Psi_{\text{Er-Sch}} = \frac{1}{\sqrt{1+K}}c B (1+K)c\frac{1}{\sqrt{1+K}}.
\end{align}
Note that in this case, $F(K)=1/\sqrt{1+K}$, and as shown in
references \cite{Erler:2009uj,Schnabl:2010tb} there is no need for
a phantom like term. It is possible to provide an integral
representation of the solution (\ref{Sol04}), this is given by
writing the inverse square root of $1 + K$ as
\begin{align}
\label{in1k} \frac{1}{\sqrt{1+K}} = \frac{1}{\sqrt{\pi}}
\int_{0}^{\infty} dt\, \frac{1}{\sqrt{t}} e^{-t} \Omega^t,
\end{align}
where $\Omega^t$ is the wedge state which can be written as \cite{Rastelli:2000iu,Schnabl:2002gg}
\begin{align}
\label{wedt} \Omega^t = e^{-Kt} = U^\dag_{\frac{2}{l}t+1}
U_{\frac{2}{l}t+1}|0 \rangle.
\end{align}

And as a last example, we consider the so-called, real tachyon
vacuum solution without square roots, or Jokel's real solution for
short \cite{Jokel:2017vlt,Arroyo:2017mpd}. This solution takes the
form
\begin{align}
\label{Soljokel} \Psi_{\text{Jok}} =
\frac{1}{4}\Big(\frac{1}{1+K}c + c\frac{1}{1+K} + c \frac{B}{1+K}
c +\frac{1}{1+K}c\frac{1}{1+K} \Big) + Q\text{-exact terms},
\end{align}
where the $Q\text{-exact terms}$ are given by
\begin{align}
\label{jokel2}  \frac{1}{2} \big[ Q(Bc)\frac{1}{1+K} +
\frac{1}{1+K}Q(Bc) \big] + \frac{1}{4}
\frac{1}{1+K}Q(Bc)\frac{1}{1+K}.
\end{align}
Interestingly, the solution does not take the factorized form
(\ref{Sol01}), and is both real and simple, namely without square
roots and phantom terms. For this real solution, the corresponding
energy has been computed and shown that the value is in agreement
with the value predicted by Sen's conjecture.

\section{The gauge invariant overlap: analytical computations}
In this section, we are going to study the analytic computation of
the gauge invariant overlap for solutions given in terms of
elements in the $KBc$ algebra. This gauge invariant observable has
been considered in references \cite{Hashimoto:2001sm,
Gaiotto:2001ji,Kawano:2008ry,Baba:2012cs,Arroyo:2019liw,Kudrna:2018mxa}.
For a given solution $\Psi$ of the string field equations of
motion, the gauge invariant overlap is defined as the evaluation
of the quantity
\begin{align}
\label{inv1} \langle \mathcal{V} | \Psi \rangle  = \langle I |
\mathcal{V}(i) | \Psi \rangle,
\end{align}
where $| I \rangle$ is the identity string field, and the operator
$\mathcal{V}(i)$ is an on-shell closed string vertex operator
$\mathcal{V}=c \tilde c V^{\text{m}}$\footnote{$V^{\text{m}}$ is a
weight $(1,1)$ conformal matter primary field.} which is inserted
at the midpoint of the string field $\Psi$. As argued by Ellwood
\cite{Ellwood:2008jh}, the gauge invariant overlap represents the
shift in the closed string tadpole of the solution relative to the
perturbative vacuum.

To evaluate the gauge invariant overlap for solutions given in
terms of elements in the $KBc$ algebra, it will be useful the
following results
\begin{align}
\label{clo1} \langle \mathcal{V} | \Omega^{t_1} c \Omega^{t_2}
\rangle &= (t_1+t_2) \mathcal{C}_{\mathcal{V}}, \\
\label{clo2} \langle \mathcal{V} | \Omega^{t_1} B  c \Omega^{t_2}
c \Omega^{t_3}
\rangle &= t_2 \, \mathcal{C}_{\mathcal{V}}, \\
\label{clo3} \langle \mathcal{V} | \Omega^{t_1} c \Omega^{t_2} B c
\Omega^{t_3} \rangle &= (t_1+t_3) \mathcal{C}_{\mathcal{V}},
\end{align}
where the coefficient $\mathcal{C}_{\mathcal{V}}$ represents the
correlator
\begin{align}
\label{clo4} \mathcal{C}_{\mathcal{V}} = \langle \mathcal{V}(i
\infty)c(0) \rangle_{C_1},
\end{align}
which is the closed string tadpole evaluate on a cylinder $C_1$ of unit circumference. The proofs of the above results (\ref{clo1})-(\ref{clo3}) are based
on usual scaling arguments and can be found in references \cite{Erler:2009uj,Murata:2011ep}.

As an application of equations (\ref{clo1})-(\ref{clo3}), we are going to compute the gauge invariant overlap for Schnabl's tachyon vacuum solution. We
would like to mention that in reference \cite{Kawano:2008ry}, after performing lengthy computations the authors have evaluated the gauge invariant overlap
for Schnabl's solution. However, as we will see, this computation can be performed in a few lines if one uses Schnabl's solution expressed in terms of the
basic string fields $K$, $B$ and $c$
\begin{eqnarray}
\label{clo5} \langle \mathcal{V}| \Psi_{\text{Sch}} \rangle = \frac{2}{l} \lim_{N\rightarrow \infty} \Big[\langle \mathcal{V} | \psi_N
\rangle-\sum_{n=0}^{N-1} \frac{d \langle \mathcal{V} | \psi_n \rangle}{dn}  \Big],
\end{eqnarray}
therefore, we need to compute $\langle \mathcal{V} | \psi_n \rangle$. Using equation (\ref{psin1}), we can write
\begin{eqnarray}
\label{clo6} \langle \mathcal{V} | \psi_n \rangle = \langle \mathcal{V}|  e^{-lK/4} cB e^{-lK n/2 } c e^{-lK/4} \rangle = \langle \mathcal{V}| \Omega^{l/4}
cB \Omega^{ln/2 } c \Omega^{l/4} \rangle.
\end{eqnarray}
Employing equation (\ref{clo3}), from equation (\ref{clo6}), we get
\begin{eqnarray}
\label{clo7} \langle \mathcal{V} | \psi_n \rangle = \frac{l}{2} \, \mathcal{C}_{\mathcal{V}},
\end{eqnarray}
plugging this result (\ref{clo7}) into equation (\ref{clo5}), we obtain
\begin{eqnarray}
\label{clo8} \langle \mathcal{V} | \Psi_{\text{Sch}} \rangle = \mathcal{C}_{\mathcal{V}} = \langle \mathcal{V}(i \infty)c(0) \rangle_{C_1}.
\end{eqnarray}
This result coincides with the expected answer of closed string
tadpole on the disk \cite{Ellwood:2008jh}. Note that the result
(\ref{clo8}) does not depend on the parameter $l$ which explicitly
appears in the solution (\ref{Sol03}).

Next we would like to evaluate the gauge invariant overlap for Erler-Schnabl's solution. Actually, using a non-real version of the solution (\ref{Sol04}),
the computation of the gauge invariant overlap has been performed in reference \cite{Erler:2009uj}. Here we are going to present the computation for the
case of the real solution\footnote{The reality condition of a string field is defined as $\Psi^{\ddag}=\Psi$ where the operation $\ddag$ means the
composition of BPZ and Hermitian conjugation. Since the basic string fields $K$, $B$ and $c$ are real string fields in this sense, the reality condition
requires that the string field read the same way from the left as from the right.}. Let us write the real solution (\ref{Sol04}) as the following integral
representation
\begin{eqnarray}
\label{cloreal1} \Psi_{\text{Er-Sch}} = \frac{1}{\pi}   \Big[ \big(1-\partial_s\big) \int_{0}^{\infty}dt_1 dt_2 \frac{e^{-t_1-t_2}}{\sqrt{t_1 t_2}}
\Omega^{t_1}cB\Omega^{s}c\Omega^{t_2}\Big]\Big{|}_{s=0},
\end{eqnarray}
therefore the gauge invariant overlap for this solution (\ref{cloreal1}) will be given by
\begin{eqnarray}
\label{cloreal2} \langle \mathcal{V}| \Psi_{\text{Er-Sch}} \rangle = \frac{1}{\pi}  \Big[ \big(1-\partial_s\big) \int_{0}^{\infty}dt_1 dt_2
\frac{e^{-t_1-t_2}}{\sqrt{t_1 t_2}} \langle \mathcal{V}| \Omega^{t_1}cB\Omega^{s}c\Omega^{t_2} \rangle \Big]\Big{|}_{s=0} .
\end{eqnarray}
Employing equation (\ref{clo3}), from equation (\ref{cloreal2}), we write
\begin{eqnarray}
\langle \mathcal{V}| \Psi_{\text{Er-Sch}} \rangle &=& \frac{1}{\pi}  \Big[ \big(1-\partial_s\big) \int_{0}^{\infty}dt_1 dt_2 \frac{e^{-t_1-t_2}}{\sqrt{t_1
t_2}} (t_1+t_2)\, \mathcal{C}_\mathcal{V} \Big]\Big{|}_{s=0} \nonumber \\
&=& \frac{1}{\pi}  \int_{0}^{\infty}dt_1 dt_2 \frac{e^{-t_1-t_2}}{\sqrt{t_1 t_2}} (t_1+t_2)\, \mathcal{C}_\mathcal{V} \nonumber \\
\label{cloreal3}  &=& \mathcal{C}_{\mathcal{V}} = \langle \mathcal{V}(i \infty)c(0) \rangle_{C_1}.
\end{eqnarray}
As we can see, this result (\ref{cloreal3}) is exactly the same as the one obtained for Schnabl's solution (\ref{clo8}).

And as the last example of analytical calculation, let us evaluate
the gauge invariant overlap for Jokel's real solution. Since BRST
exact terms do not contribute for the evaluation of the gauge
invariant overlap, we just need to consider the non-BRST exact
terms of the solution. These terms are given on the right hand
side of equation (\ref{Soljokel}) and they can be written as
\begin{align}
 \hat{\Psi}_{\text{Jok}} &\equiv
\frac{1}{4}\Big(\frac{1}{1+K}c + c\frac{1}{1+K} + c \frac{B}{1+K}
c +\frac{1}{1+K}c\frac{1}{1+K} \Big) \nonumber \\
\label{Soljokelg1} &=\frac{1}{4} \int_{0}^{\infty}dt \,
e^{-t}\,(\Omega^t c+c \Omega^t + c  \Omega^t B c) +
\frac{1}{4}\int_{0}^{\infty}dt_1 dt_2 \, e^{-t_1-t_2} \,
\Omega^{t_1} c \Omega^{t_2}.
\end{align}
Therefore the gauge invariant overlap for Jokel's real solution is
given by
\begin{align}
\label{Soljokelg2} \langle \mathcal{V}| \hat{\Psi}_{\text{Jok}}
\rangle = \frac{1}{4} \int_{0}^{\infty}dt \, e^{-t}\,\langle
\mathcal{V}|\Omega^t c+c \Omega^t + c  \Omega^t B c\rangle +
\frac{1}{4}\int_{0}^{\infty}dt_1 dt_2 \, e^{-t_1-t_2} \, \langle
\mathcal{V}|\Omega^{t_1} c \Omega^{t_2}\rangle.
\end{align}
Using equations (\ref{clo1}) and (\ref{clo3}), from equation
(\ref{Soljokelg2}) we obtain
\begin{align}
 \langle \mathcal{V}| \hat{\Psi}_{\text{Jok}}
\rangle &= \Big[\frac{1}{2} \int_{0}^{\infty}dt \; t e^{-t}
 + \frac{1}{4}\int_{0}^{\infty}dt_1 dt_2
\, (t_1+t_2) e^{-t_1-t_2} \Big]  \mathcal{C}_{\mathcal{V}}
\nonumber \\
\label{Soljokelg3} &= \mathcal{C}_{\mathcal{V}} = \langle
\mathcal{V}(i \infty)c(0) \rangle_{C_1}.
\end{align}
Note that this result (\ref{Soljokelg3}) is the same as the ones
obtained in the case of Schnabl's (\ref{clo8}) and Erler-Schnabl's
solutions (\ref{cloreal3}).

It should be nice to obtain the above analytic results by
numerical means. For instance, using the traditional Virasoro
$L_0$ level truncation scheme, in reference \cite{Kawano:2008ry},
the authors have evaluated the gauge invariant overlap for
Schnabl's solution. However, up to now, using the Virasoro $L_0$
level truncation scheme, the analysis of the gauge invariant
overlap for Erler-Schnabl's and Jokel's real solution was not
performed. Moreover, the analysis of the gauge invariant overlap
by means of the curly $\mathcal{L}_0$ level truncation scheme has
not been carried out neither for Schnabl's, Erler-Schnabl's nor
for Jokel's real solution.

In the next two sections, using the curly $\mathcal{L}_0$ and the
Virasoro $L_0$ level truncation scheme, we are going to present
the evaluation of the gauge invariant overlap for solutions
constructed out of elements in the $KBc$ algebra.

\section{The gauge invariant overlap: $\mathcal{L}_0 $ level truncation computations}

Since from the beginning, we do not know if the result for the
gauge invariant overlap obtained by analytical computations will
match the result obtained by numerical means (either by using the
$\mathcal{L}_0$ or the $L_0$ level truncation scheme), it is
important for the consistency of the solutions to explicitly check
if these different schemes provide the same answer. In this
section, using the $\mathcal{L}_0$ level expansion of a rather
generic solution $\Psi$, we will present the evaluation of the
gauge invariant overlap.

As we know, the solution is given in terms of elements in the
$KBc$ algebra (which involves the operators $\hat{\mathcal{L}}$,
$\hat{\mathcal{B}}$ and $\tilde c$), in general, we can write the
following $\mathcal{L}_0$ level expansion
\begin{eqnarray}
\label{psicurl0} \Psi = \sum_{n,p} f_{n,p} \hat{\mathcal{L}}^{n}
\tilde c_p |0\rangle + \sum_{n,p,q} f_{n,p,q}
\hat{\mathcal{L}}^{n} \hat{\mathcal{B}} \tilde c_p \tilde c_q
|0\rangle ,
\end{eqnarray}
where $n = 0 , 1 , 2 , \cdots$, and $p, q = 1 , 0 , - 1 , - 2 , \cdots$. The coefficients of the expansion $f_{n,p}$ and $f_{n,p,q}$ can be regarded as
generic ones, obviously these coefficients depend on the solution we choose. For instance, for the case of Schnabl's solution (\ref{Sol02}), these
coefficients are given by
\begin{eqnarray}
\label{coeffsch1} f_{n, p} &=&
\frac{1-(-1)^p}{2}\frac{l^{-p}}{2^{n - 2 p + 1}} \frac{1}{n!}
(-1)^n  B_{n - p + 1} \, , \\
\label{coeffsch2} f_{n, p, q} &=& \frac{1-(-1)^{p + q}}{2}
\frac{l^{-p - q}}{2^{n - 2 (p + q) + 3}} \frac{1}{n!} (-1)^{n -q}
B_{n - p - q + 2} \, ,
\end{eqnarray}
where $B_m$ are the Bernoulli's numbers.

To compute the gauge invariant overlap for solutions expanded in
terms of $\mathcal{L}_0$ eigenstates, we start by replacing the
string field $\Psi$ with $z^{\mathcal{L}_0} \Psi$, so that states
in the $\mathcal{L}_0$ level expansion will acquire different
integer powers of $z$ at different levels. And as usual, at the
end, we will simply set $z = 1$.

Let us start with the evaluation of the gauge invariant overlap as a formal power series expansion in $z$. Plugging the expansion (\ref{psicurl0}) into the
definition of the gauge invariant overlap (\ref{inv1}), we obtain
\begin{align}
\label{invcurly1} \langle \mathcal{V} | z^{\mathcal{L}_0} \Psi
\rangle = \sum_{n,p} z^{n-p}f_{n,p} \langle \mathcal{V}
|\hat{\mathcal{L}}^{n} \tilde c_p |0\rangle + \sum_{n,p,q}
z^{n+1-p-q} f_{n,p,q} \langle \mathcal{V} |\hat{\mathcal{L}}^{n}
\hat{\mathcal{B}} \tilde c_p \tilde c_q |0\rangle.
\end{align}
As we can see, we need to compute $\langle \mathcal{V}
|\hat{\mathcal{L}}^{n} \tilde c_p |0\rangle$ and $\langle
\mathcal{V} |\hat{\mathcal{L}}^{n} \hat{\mathcal{B}} \tilde c_p
\tilde c_q |0\rangle$. To evaluate these quantities, we require
express $\hat{\mathcal{L}}^{n} \tilde c_p |0\rangle$ and
$\hat{\mathcal{L}}^{n} \hat{\mathcal{B}} \tilde c_p \tilde c_q
|0\rangle$ in terms of elements in the $KBc$ algebra, for this
purpose, it will be useful the following relations
\begin{align}
\label{relclo1} \Omega^{s_1} c \Omega^{s_2} &= e^{u
\hat{\mathcal{L}}} \tilde c(x) |0\rangle, \\
\label{relclo2} B\Omega^{t_1} c \Omega^{t_2} c \Omega^{t_3} -
\frac{1}{2} \Omega^{t_1+t_2} c \Omega^{t_3}+\frac{1}{2}
\Omega^{t_1} c \Omega^{t_2+t_3} &= \frac{1}{l}\hat{\mathcal{B}}
e^{u \hat{\mathcal{L}}} \tilde c(x)\tilde c(y) |0\rangle,
\end{align}
where
\begin{align}
\label{relclo3} s_1 = \frac{l}{4}-\frac{l u}{2} - x , \;\; s_2 =
\frac{l}{4}-\frac{l u}{2} + x, \\
\label{relclo4} t_1 = \frac{l}{4}-\frac{l u}{2} - x , \;\; t_2 =
x-y , \;\; t_3 = \frac{l}{4}-\frac{l u}{2} + y.
\end{align}
Employing the above relations, we can write $\hat{\mathcal{L}}^{n}
\tilde c_p |0\rangle$ and $\hat{\mathcal{L}}^{n} \hat{\mathcal{B}}
\tilde c_p \tilde c_q |0\rangle$ in terms of elements in the $KBc$
algebra
\begin{align}
\label{relclo5} \hat{\mathcal{L}}^n \tilde c_p |0\rangle &= n!
\oint \frac{du}{2 \pi i} \frac{dx}{2 \pi i} u^{-n-1} x^{p-2}
  \Omega^{s_1} c \Omega^{s_2}, \\
\label{relclo6} \hat{\mathcal{L}}^{n} \hat{\mathcal{B}} \tilde c_p
\tilde c_q |0\rangle &= n! \oint \frac{du}{2 \pi i} \frac{dx}{2
\pi i} \frac{dy}{2 \pi i} u^{-n-1} x^{p-2} y^{q-2} \big[ l
B\Omega^{t_1} c \Omega^{t_2} c \Omega^{t_3} - \frac{l}{2}
\Omega^{t_1+t_2} c \Omega^{t_3}+\frac{l}{2} \Omega^{t_1} c
\Omega^{t_2+t_3} \big].
\end{align}

Now we are in position to evaluate the quantities $\langle
\mathcal{V} |\hat{\mathcal{L}}^{n} \tilde c_p |0\rangle$ and
$\langle \mathcal{V} |\hat{\mathcal{L}}^{n} \hat{\mathcal{B}}
\tilde c_p \tilde c_q |0\rangle$. For instance, using equations
(\ref{clo1}) and (\ref{relclo5}), let us compute
\begin{align}
\langle \mathcal{V} | \hat{\mathcal{L}}^n \tilde c_p |0\rangle &=
n! \oint \frac{du}{2 \pi i} \frac{dx}{2 \pi i} u^{-n-1} x^{p-2}
\langle \mathcal{V} |  \Omega^{s_1} c \Omega^{s_2} \rangle
\nonumber \\
&= n! \oint \frac{du}{2 \pi i} \frac{dx}{2 \pi i} u^{-n-1} x^{p-2}
\big(\frac{l}{2}-l u\big) \mathcal{C}_\mathcal{V} \nonumber \\
\label{relclo7} &= l \, \big(\frac{\delta_{p,1} \delta_{n,0}}{2}-
\delta_{p,1} \delta_{n,1}\big)\mathcal{C}_\mathcal{V}.
\end{align}
Performing similar calculations as the above, using equations
(\ref{clo1}), (\ref{clo2}) and (\ref{relclo6}), we obtain
\begin{align}
\label{relclo8} \langle \mathcal{V} | \hat{\mathcal{B}}
\hat{\mathcal{L}}^n \tilde c_p \tilde c_q |0\rangle = l \,
\big(\delta_{n,0}\delta_{p,0}\delta_{q,1}-\delta_{n,0}\delta_{q,0}\delta_{p,1}\big)\mathcal{C}_\mathcal{V}.
\end{align}

Finally, plugging the results (\ref{relclo7}) and (\ref{relclo8})
into the definition of the gauge invariant overlap
(\ref{invcurly1}), and setting $z=1$, we get
\begin{align}
\label{relclo9} \langle \mathcal{V} | \Psi \rangle = l \,
\big(\frac{f_{0,1}}{2}-f_{1,1}-2f_{0,1,0}\big)
\mathcal{C}_\mathcal{V}.
\end{align}
To compute the gauge invariant overlap for a solution expanded in terms of $\mathcal{L}_0$ eigenstates (\ref{psicurl0}), we only need to know the value of
the first three coefficients appearing at levels $z^{-1}$ and $z^{0}$. Remarkably, this result (\ref{relclo9}) is simpler than the one obtained for the
case of the energy. Evaluating the energy in the $\mathcal{L}_0$ level expansion gives a very complicated  non-convergent series, though the series can be
resummed numerically by means of the so-called Pad\'{e} approximants to give a good approximation to the brane tension
\cite{Schnabl:2005gv,Erler:2009uj,Arroyo:2009ec}.

Let us apply the general result (\ref{relclo9}) for some particular solutions such as the Schnabl's solution $\Psi_{\text{Sch}}$. Using the explicit
expressions of the coefficients (\ref{coeffsch1}) and (\ref{coeffsch2})
\begin{align}
\label{relclo10} f_{0,1}=\frac{2}{l}, \;\;\; f_{1,1}=\frac{1}{2l}, \;\;\; f_{0,1,0}=-\frac{1}{4l},
\end{align}
which appear in the $\mathcal{L}_0$ level expansion of Schnabl's solution, from equation (\ref{relclo9}) we obtain
\begin{align}
\label{relclo11} \langle \mathcal{V} | \Psi_{\text{Sch}} \rangle =  \mathcal{C}_\mathcal{V}.
\end{align}
This result does not depend on the parameter $l$ and is the same result as the one obtained from analytic computations.

In the case of Erler-Schnabl's solution $\Psi_{\text{Er-Sch}}$, using its integral representation (\ref{cloreal1}), we can compute the first three
coefficients appearing in the $\mathcal{L}_0$ level expansion of the solution
\begin{align}
\label{relclo12} f_{0,1}=1, \;\;\; f_{1,1}=\frac{1}{2}, \;\;\; f_{0,1,0}=-\frac{1}{2l}.
\end{align}
Therefore, plugging these results (\ref{relclo12}) into equation (\ref{relclo9}), we obtain
\begin{align}
\label{relclo13} \langle \mathcal{V} | \Psi_{\text{Er-Sch}} \rangle =  \mathcal{C}_\mathcal{V}.
\end{align}
This result also does not depend on the parameter $l$ and is the same result as the one obtained for the case of Schnabl's solution.

In the case of Jokel's real solution, we can also calculate the
curly $\mathcal{L}_0$ level expansion of the non-BRST exact terms
of the solution (\ref{Soljokelg1}). The first three coefficients
of this $\mathcal{L}_0$ level expansion are given by
\begin{align}
\label{jokcoef1} f_{0,1}=\frac{2}{l}, \;\;\;
f_{1,1}=-\frac{1}{4l}, \;\;\; f_{0,1,0}=\frac{1}{8l}.
\end{align}
Substituting these results (\ref{jokcoef1}) into equation
(\ref{relclo9}), we get
\begin{align}
\label{joktadpolel0} \langle \mathcal{V}| \hat{\Psi}_{\text{Jok}}
\rangle = \mathcal{C}_\mathcal{V}.
\end{align}
As we can see, the result (\ref{joktadpolel0}) is the same as the
ones obtained in the case of Schnabl's and Erler-Schnabl's
solutions.

So far, we have computed the gauge invariant overlap by two means: analytically and using the curly $\mathcal{L}_0$ level expansion of the solutions. In
what follows, we are going to evaluate the gauge invariant overlap by a third method, namely, using the traditional Virasoro $L_0$ level expansion of the
solutions.

\section{The gauge invariant overlap: Virasoro $L_0 $ level truncation computations}
In this section, using the $L_0$ level truncation scheme, the evaluation of the gauge invariant overlap will be shown. Since the solution $\Psi$ involves
the operators $\hat{\mathcal{L}}$, $\hat{\mathcal{B}}$ and $\tilde c$, we can write its $L_0$ level expansion as follows
\begin{eqnarray}
\label{psiL01} \Psi = \sum g_{n_1 n_2\cdots n_i p} L_{n_1}
L_{n_2}\cdots L_{n_i}c_p |0\rangle + \sum g_{m_1 m_2\cdots m_j s p
q} L_{m_1} L_{m_2}\cdots L_{m_j} b_s c_p c_q |0\rangle ,
\end{eqnarray}
where $n_i,m_j,s \leq -2$ and $p,q=1,0,-1,-2,\cdots$. The $L_n$'s are the ordinary Virasoro generators with zero central charge $c=0$ of the total (i.e.
matter and ghost) conformal field theory. For instance, Schnabl's solution (\ref{Sol03}), with $l=\pi$, expanded up to level two states is given by
\begin{eqnarray}
\label{psiL02} \Psi_{\text{Sch}} =  0.553465 \, c_1|0\rangle + 0.043671 \, c_{-1}|0\rangle +0.137646 \, L_{-2}c_1|0\rangle + 0.131082 \, b_{-2} c_0 c_1
|0\rangle .
\end{eqnarray}

To compute the gauge invariant overlap by means of the $L_0$ level truncation scheme, it is clear that if we insert the expansion (\ref{psiL01}) into the
definition of the gauge invariant overlap (\ref{inv1}), we will need to evaluate the quantities
\begin{eqnarray}
\label{psiL03}  \langle \mathcal{V} | L_{n_1} L_{n_2}\cdots
L_{n_i}c_p |0\rangle, \;\;\;\;   \langle \mathcal{V} | L_{m_1}
L_{m_2}\cdots L_{m_j} b_s c_p c_q |0\rangle.
\end{eqnarray}
We are going to calculate these quantities by means of a recursive method based on the evaluation of the following commutation and anti-commutation
relations
\begin{align}
\label{psiL04} [L_m,L_n]&= (m-n) L_{m+n}, \\
\label{psiL05} [L_m,b_n]&= (m-n)  b_{m+n},  \\
\label{psiL06} [L_n, c_p]&= (-2n-p)   c_{n+p}, \\
\label{psiL07} \{b_m, c_n\} &= \delta_{m+n,0}.
\end{align}

As an illustration, suppose we need to calculate $\langle \mathcal{V} | L_{n} c_p |0\rangle$. Since for $n \leq -2$ the operator $L_n$ does not annihilate
the vacuum $|0\rangle$, and in order to apply the commutator (\ref{psiL06}), we must first express the operator $L_n$ in terms of annihilation operators.
This can be achieved if we use the fact that the on-shell closed string state $\mathcal{V}=c \tilde c V^{\text{m}}$ is invariant by the transformation
generated by $K_n = L_n -(-1)^n L_{-n}$, namely, we have \cite{Kawano:2008ry}
\begin{align}
\label{psiL08}  \langle \mathcal{V} | L_n =  \langle \mathcal{V}
|(-1)^n L_{-n}.
\end{align}
And now, since $L_{-n}|0\rangle=0$ for $n \leq -2$,  we are able
to compute $\langle \mathcal{V} | L_{n} c_p |0\rangle$ using the
commutator (\ref{psiL06})
\begin{align}
\label{psiL09}  \langle \mathcal{V} | L_n c_p |0\rangle = (-1)^n
\langle \mathcal{V} | [L_{-n},c_p] |0\rangle = (-1)^n (2n-p)
\langle \mathcal{V} | c_{p-n} |0\rangle .
\end{align}
Let us comment that for the case of the operator $b_n$ which
corresponds to the modes of the ghost field $b$, we have a similar
result as the one given by equation (\ref{psiL08})
\cite{Hashimoto:2001sm,
Gaiotto:2001ji,Kawano:2008ry,Kudrna:2012re}
\begin{align}
\label{bbpsiL08}  \langle \mathcal{V} | b_n =  \langle \mathcal{V}
|(-1)^n b_{-n}.
\end{align}

As we have seen, after the use of the commutation and anti-commutation relations (\ref{psiL04})-(\ref{psiL07}), we can express the quantities
(\ref{psiL03}) as linear combinations of terms like
\begin{align}
\label{psiL010} \langle \mathcal{V} | c_p |0\rangle.
\end{align}
To evaluate (\ref{psiL010}), first let us express the mode $c_p$ in the $\tilde z$-coordinate. Using the conformal transformation of the $c(z)$ ghost,
under the map (\ref{map3}), we get
\begin{align}
\label{psiL011} c_p = \oint \frac{d z}{2\pi i} z^{p-2} c(z) = \Big(\frac{\pi}{l}\Big)^2 \oint \frac{d \tilde z}{2\pi i} \sec ^4\left(\frac{\pi  \tilde
z}{l}\right) \tan ^{p-2}\left(\frac{\pi \tilde z}{l}\right) \tilde c (\tilde z).
\end{align}
If we substitute equation (\ref{psiL011}) into equation (\ref{psiL010}), it is clear that we will need to evaluate the quantity $\langle \mathcal{V} |
\tilde c (\tilde z) |0\rangle$. Using equations (\ref{clo1}) and (\ref{relclo1}), we can compute this quantity
\begin{align}
\label{psiL012} \langle \mathcal{V} | \tilde c (\tilde z) |0\rangle = \langle \mathcal{V} | \Omega^{-\tilde z +l/4} c \, \Omega^{\tilde z +l/4} \rangle =
\frac{l}{2} \, \mathcal{C}_\mathcal{V} .
\end{align}
Therefore, employing equations (\ref{psiL011}) and (\ref{psiL012}), we obtain
\begin{align}
\label{psiL013} \langle \mathcal{V} | c_p |0\rangle = \Big(\frac{\pi}{l}\Big)^2 \Big(\frac{l}{2}\Big) \mathcal{C}_\mathcal{V} \oint \frac{d \tilde z}{2\pi
i} \sec ^4\left(\frac{\pi  \tilde z}{l}\right) \tan ^{p-2}\left(\frac{\pi \tilde z}{l}\right) =
\frac{\pi}{2}\big(\delta_{p,-1}+\delta_{p,1}\big)\,\mathcal{C}_\mathcal{V}.
\end{align}

As a first example, let us compute the gauge invariant overlap for Schnabl's solution expanded up to level two states
\begin{eqnarray}
\label{psiL14} \Psi_{\text{Sch}} =  t' \, c_1|0\rangle + u' \, c_{-1}|0\rangle +v' \, L_{-2}c_1|0\rangle + w' \, b_{-2} c_0 c_1 |0\rangle ,
\end{eqnarray}
where the values of the coefficients $t'$, $u'$, $v'$ and $w'$ are given in equation (\ref{psiL02}). Using the property that $\langle \mathcal{V} | L_{-2}
= \langle \mathcal{V} |L_{2}$ and $\langle \mathcal{V} | b_{-2} = \langle \mathcal{V} |b_{2}$, the evaluation of the gauge invariant overlap reads as
follows
\begin{align}
\langle \mathcal{V} | \Psi_{\text{Sch}} \rangle &=  t' \langle \mathcal{V} | c_1|0\rangle + u' \langle \mathcal{V} | c_{-1}|0\rangle + v' \langle
\mathcal{V} | [L_{2},c_1]|0\rangle + w' \langle \mathcal{V} |
[b_{2}, c_0 c_1] |0\rangle \nonumber \\
\label{psiL15} &=t' \langle \mathcal{V} | c_1|0\rangle + u' \langle \mathcal{V} | c_{-1}|0\rangle -5 v' \langle \mathcal{V} | c_3 |0\rangle =
\frac{\pi}{2}\big(t'+u'\big)\,\mathcal{C}_\mathcal{V}.
\end{align}

We would like to compare this result (\ref{psiL15}) with the one obtained in reference \cite{Kawano:2008ry}, where Schnabl's solution has been expanded in
a slightly different basis. Instead of considering the Virasoro generators $L_n$ with zero central charge, the authors have used the $\alpha_n$'s
oscillators, for instance, up to level two states, they have written the expansion
\begin{eqnarray}
\label{psiL16} \Psi_{\text{Sch}} =  t \, c_1|0\rangle + u \, c_{-1}|0\rangle +v \,(\alpha_{-1}\cdot\alpha_{-1})c_1|0\rangle + w \, b_{-2} c_0 c_1 |0\rangle
,
\end{eqnarray}
where the coefficients have the following values\footnote{We have noted that if we use equation (3.36) of reference \cite{Kawano:2008ry}, the value of the
coefficient $v$ turns out to be twice the value presented here (\ref{psiL17}). This means that if the authors want to use the definition of $v$ as given in
their equations (3.31) and (3.32), their equation (3.36) should be replaced by a half of it. We have communicated this issue to one of the authors, and he
has confirmed this little mistake which nevertheless does not change the main result presented in \cite{Kawano:2008ry}.}
\begin{eqnarray}
\label{psiL17} t = 0.553465, \;\; u = 0.456611, \;\; v= 0.068823, \;\; w=-0.144210 .
\end{eqnarray}

Then by using an explicit oscillator representation for the on-shell closed string state, which can be found in references
\cite{Kawano:2008ry,Takahashi:2003kq}, the gauge invariant overlap for the expanded Schnabl's solution (\ref{psiL16}) turns out to be \cite{Kawano:2008ry}
\begin{eqnarray}
\label{psiL18} \langle \mathcal{V} | \Psi_{\text{Sch}} \rangle = \frac{1}{4} t - \frac{3}{2}v + \frac{1}{4} u = 0.149284.
\end{eqnarray}
Let us compare this result (\ref{psiL18}) with the one obtained by us (\ref{psiL15}). To get the same answer, we should choose the normalization where
$\mathcal{C}_\mathcal{V} = 1/(2 \pi)$, and in fact with this normalization from equation (\ref{psiL15}), we obtain
\begin{align}
\label{psiL19} \langle \mathcal{V} | \Psi_{\text{Sch}} \rangle = \frac{1}{4}\big(t'+u'\big) = 0.149284.
\end{align}
Taking into account higher level states, we have performed the
computation of the gauge invariant overlap for Schnabl's solution,
and the results we have obtained with the normalization
$\mathcal{C}_\mathcal{V} = 1/(2 \pi)$ are in agreement with the
ones presented in reference \cite{Kawano:2008ry}. We can consider
this agreement as a test for the method of computing the gauge
invariant overlap based on the use of the equations
(\ref{psiL08}), (\ref{bbpsiL08}) and the commutation and
anti-commutation relations (\ref{psiL04})-(\ref{psiL07}).

The advantage of this method compare to the one presented in
\cite{Kawano:2008ry}, is that we do not require to use an explicit
oscillator representation for the on-shell closed string state.
The implication of this observation will be reflected in the
simplification of the evaluation of the gauge invariant overlap.
Recall that the $L_0$ level expansion of analytic solutions
constructed out of elements in the $KBc$ algebra, as presented in
(\ref{psiL01}), is naively given in terms of the total
(matter+ghost) Virasoro generators $L_n$, the $b_n$ and $c_p$
modes, and since we do not require to use an explicit oscillator
representation for the on-shell closed string state, using the
expansion (\ref{psiL01}) we can directly evaluate the gauge
invariant overlap without the necessity of reexpressing the
expansion in terms of the $\alpha_n$'s oscillators (which will
require an additional work).

Before to study the numerical evaluation of the gauge invariant
overlap for the case of Erler-Schnabl's and Jokel's solutions, we
would like to mention some motivations for doing this computation.
Firstly, using the $L_0$ level truncation scheme, the numerical
analysis of the gauge invariant overlap for Erler-Schnabl's and
Jokel's solutions has not been carried out. This analysis should
be crucial if we want to confirm the analytic result. However, the
main motivation for performing such numerical computations is to
see whether or not higher level contributions yield to
increasingly convergent results which approach to the expected
answer. In the case of Schnabl's solution, it has been shown that
every time we increase the level of the truncated solution, the
gauge invariant overlap converges to the expected analytical
result without the necessity of using any regularization scheme
such as Pad\'{e} approximants \cite{Kawano:2008ry}.

Let us start with the $L_0$ level truncation analysis of the gauge invariant overlap for Erler-Schnabl's solution. To simplify the computations, it will be
useful to write the solution (\ref{Sol04}) in the following way
\begin{eqnarray}
\label{PsiBrst} \Psi_{\text{Er-Sch}} = \frac{1}{\sqrt{1+ K}} c \frac{1}{\sqrt{1+K}} + Q\Big\{ \frac{1}{\sqrt{1+K}} Bc \frac{1}{\sqrt{1+K}} \Big\}.
\end{eqnarray}
Inserting the solution (\ref{PsiBrst}) into the definition of the gauge invariant overlap, the BRST exact term does not contribute, and so we only need to
consider the first term appearing on the right hand side of equation (\ref{PsiBrst}), let us denote this term as
\begin{eqnarray}
\label{PsiBrsthat} \Psi^{(1)} \equiv \frac{1}{\sqrt{1+ K}} c \frac{1}{\sqrt{1+K}}.
\end{eqnarray}

To compare the $L_0$ level expansion of the string field (\ref{PsiBrsthat}) with the one presented in reference \cite{Erler:2009uj}, we choose the value of
the parameter $l$, which appears in the definition of the map (\ref{map3}), as $l=2$. The $L_0$ level expansion of the string field (\ref{PsiBrsthat}) can
be obtained from the following result \cite{Erler:2009uj,Arroyo:2014pua}
\begin{eqnarray}
\label{psiFirst1} \Psi^{(1)} = \frac{1}{2 \pi^2 } \int_{0}^{\infty}ds dt \, \frac{1}{\sqrt{st}} e^{-s-t} r^2 \cos ^2\left(\frac{\pi x}{r}\right)
\widetilde{U}_{r} c\left(\frac{2 \tan \left(\frac{\pi
   x}{r}\right)}{r}\right)|0\rangle ,
\end{eqnarray}
where $r$ and $x$ are given by
\begin{eqnarray}
\label{rx} r=s+t+1 , \;\;\;\;\; x=\frac{s-t}{2}.
\end{eqnarray}
The operator $\widetilde{U}_{r}$ is defined as
\begin{eqnarray}
\label{uux2} \widetilde{U}_{r} \equiv  \cdots e^{u_{10,r} L_{-10}} e^{u_{8,r} L_{-8}} e^{u_{6,r} L_{-6}} e^{u_{4,r} L_{-4}}e^{u_{2,r} L_{-2}}.
\end{eqnarray}
To find the coefficients $u_{n,r}$ appearing in the exponentials, we use
\begin{align}
\frac{r}{2} \tan (\frac{2}{r} \arctan z) &= \lim_{N \rightarrow \infty} \big[f_{2,u_{2,r}} \circ  f_{4,u_{4,r}} \circ f_{6,u_{6,r}} \circ f_{8,u_{8,r}}
\circ
f_{10,u_{10,r}} \circ \cdots \circ f_{N,u_{N,r}}(z)\big] \nonumber \\
&= \lim_{N \rightarrow \infty}\big[ f_{2,u_{2,r}} ( f_{4,u_{4,r}} ( f_{6,u_{6,r}} ( f_{8,u_{8,r}} ( f_{10,u_{10,r}}(\cdots (f_{N,u_{N,r}}(z)) \dots )))))
\big],
\end{align}
where the function $f_{n,u_{n,r}}(z)$ is given by
\begin{eqnarray}
f_{n,u_{n,r}}(z) = \frac{z}{(1-u_{n,r} n z^n)^{1/n}}.
\end{eqnarray}

By performing the change of variables
\begin{eqnarray}
s\to \frac{1}{2} (u-u \eta), \;\;\;\; t\to \frac{1}{2} (u+u \eta) , \;\;\;\; ds dt \to \frac{u}{2} dud\eta,
\end{eqnarray}
where $u \in[0,\infty)$ and $\eta \in (-1,1)$, we are going to numerically evaluate the double integrals coming from equation (\ref{psiFirst1}).

Employing the above results, let us write the string field
(\ref{PsiBrsthat}), expanded up to level four states
\begin{align}
\Psi^{(1)} =&+0.509038 \, c_1|0\rangle + 0.13231 \, c_{-1}|0\rangle -0.001576 \, L_{-2}c_1|0\rangle + 0.0893356 \, c_{-3} |0\rangle
\nonumber \\
\label{psiErSch1} &-0.0135795 \, L_{-4}c_1|0\rangle -0.00694698 \,
L_{-2} c_{-1}|0\rangle + 0.0231579 \,L_{-2} L_{-2}c_1|0\rangle .
\end{align}

To evaluate the gauge invariant overlap using the $L_0$ level truncation scheme, first we perform the replacement $\Psi^{(1)} \rightarrow z^{L_0}
\Psi^{(1)}$ and then using the resulting string field $z^{L_0} \Psi^{(1)}$, we define
\begin{eqnarray}
\label{gaugedependz} \langle \mathcal{V} | \Psi^{(1)} \rangle(z) \equiv  \langle \mathcal{V} | z^{L_0} \Psi^{(1)} \rangle.
\end{eqnarray}
The value of the gauge invariant overlap is obtained just by setting $z = 1$. As we can see, our problem has been reduced to the computation of quantities
like $\langle \mathcal{V} |L_{n_1} L_{n_2}\cdots L_{n_i}c_p |0\rangle$ which can be evaluated using equations (\ref{psiL04})-(\ref{psiL07}), (\ref{psiL08})
and (\ref{psiL013}).

As an example, plugging the level expansion (\ref{psiErSch1}) into the definition (\ref{gaugedependz}), we obtain
\begin{align}
\label{gaugeexpan1}   \langle \mathcal{V} | \Psi^{(1)} \rangle(z)
= \Big[ \frac{0.79959514}{z} + 0.20783242 z - 0.11276868 z^3 \Big]
\mathcal{C}_\mathcal{V}.
\end{align}
If we set $z = 1$, from equation (\ref{gaugeexpan1}) we get about $89\%$ of the expected result for the gauge invariant overlap (\ref{cloreal3}). This
result may appear good, however, considering the string field (\ref{PsiBrsthat}) expanded up to level twenty four states, we obtain about $116\%$ of the
expected result. This behavior is in contrast with the case of Schnabl's solution, where it has been shown that every time we increase the level of the
truncated solution, the gauge invariant overlap converges to the expected analytical result \cite{Kawano:2008ry}. Therefore, as we suspect, for the case of
Erler-Schnabl's solution, by naively setting $z=1$, we are obtaining a non-convergent result. Recall that in numerical $L_0$ level truncation computations
a regularization procedure based on Pad\'{e} approximants produces desired results for gauge invariant quantities like the energy \cite{Erler:2009uj}. Let
us see if after applying Pad\'{e} approximants, we can obtain the expected answer for the case of the gauge invariant overlap.

To obtain a Pad\'{e} approximant of order $P^n_{1+n}(z)$ for the
gauge invariant overlap, we will need to know the series expansion
of (\ref{gaugedependz}) up to the order $z^{2n-1}$. For the
numerical evaluation, we have considered the string field
$\Psi^{(1)}$ expanded up to level twenty four states, so that we
obtain a series expansion for (\ref{gaugedependz}) truncated up to
the order $z^{23}$. The explicit expression for the gauge
invariant overlap, truncated up to this order, is given by
\begin{align}
\langle \mathcal{V} | \Psi^{(1)} \rangle(z) = \Big[ & \frac{0.79959514}{z} + 0.20783242 z - 0.11276868 z^3 +
 0.03183002 z^5 \nonumber \\
&+0.1105491863 z^7 + 0.003197445654 z^9 -
 0.14509620056 z^{11} \nonumber \\
&+0.0040708415 z^{13} + 0.1939886423 z^{15} +
 0.002321956902 z^{17} \nonumber \\
 \label{gaugeexpanlevel11} &-0.2468785966 z^{19} + 0.0009635172 z^{21} +
 0.313942988469 z^{23} \Big] \mathcal{C}_\mathcal{V}.
\end{align}

As an illustration of the numerical method based on Pad\'{e} approximants, let us compute the value of the gauge invariant overlap using a Pad\'{e}
approximant of order $P^4_{1+4}(z)$. First, we express $\langle \mathcal{V} | \Psi^{(1)} \rangle(z)$ as the rational function $P^4_{1+4}(z)$
\begin{eqnarray}
\label{ss2} \langle \mathcal{V} | \Psi^{(1)} \rangle(z)=P^4_{1+4}(z)=\frac{1}{z} \Big[\frac{a_0+a_1z+a_2z^2+a_3z^3+a_4z^4 }{1+b_1z+b_2z^2+b_3z^3+b_4z^4}
\Big] \mathcal{C}_\mathcal{V} \, .
\end{eqnarray}
Expanding the right hand side of (\ref{ss2}) around $z=0$ up to seventh order in $z$ and equating the coefficients of $z^{-1}$, $z^{0}$, $z^{1}$, $z^{2}$,
$z^{3}$, $z^{4}$, $z^{5}$, $z^{6}$, $z^{7}$ with the expansion (\ref{gaugeexpanlevel11}), we get a system of algebraic equations for the unknown
coefficients $a_0$, $a_1$, $a_2$, $a_3$, $a_4$, $b_1$, $b_2$, $b_3$ and $b_4$. Solving those equations we get
\begin{eqnarray}
a_0 = 0.799595, \;\;\; a_1=0, \;\;\; a_2=3.68919 ,
\;\;\; a_3=0, \;\;\; a_4=2.55861, \;\;\;\; \\
b_1 = 0, \;\;\; b_2=4.35389, \;\;\; b_3=0, \;\;\; b_4=2.20925. \;\;\;\;\;\;\;\;\;\;\;\;\;\;\;\;\;\;\;\;\;
\end{eqnarray}
Replacing the value of these coefficients inside the definition of $P^4_{1+4}(z)$ (\ref{ss2}), and evaluating this at $z=1$, we get the following value of
the gauge invariant overlap
\begin{eqnarray}
\label{gauge01x} P^4_{1+4}(z=1) = 0.931807965 \; \mathcal{C}_\mathcal{V}.
\end{eqnarray}

The results of our calculations are summarized in table \ref{realresultsx1}. As we can see, the value of the gauge invariant overlap evaluated using
Pad\'{e} approximants confirms the expected analytic result (\ref{cloreal3}). Although the convergence to the expected answer gets quite slow, by
considering higher level contributions, we will eventually reach to the right value of the gauge invariant overlap $\langle \mathcal{V} | \Psi^{(1)}
\rangle \rightarrow 1 \, \mathcal{C}_\mathcal{V}$.

\begin{table}[ht]
\caption{The Pad\'{e} approximation for the value of the gauge invariant overlap $\langle \mathcal{V} | z^{L_0} \Psi^{(1)} \rangle$ divided by
$\mathcal{C}_\mathcal{V}$ and evaluated at $z=1$. The third column shows the $P_{1+n}^{n}$ Pad\'{e} approximation. In the last column, $P^{2n}_1$
represents a trivial approximation, a naively summed series. At each line, we have considered the string field expanded up to level $2n$ states.}
\centering
\begin{tabular}{|c|c|c|c|}
\hline
  & Level  & $P^{n}_{1+n}$ &  $P^{2n}_1$   \\
    \hline $n=0$ & 0 & $0.7995951404$ & $0.7995951404$ \\
\hline  $n=2$& 4 & $0.9343242915$  & $0.8946588687$ \\
\hline $n=4$ & 8 & $0.9318079653$  & $1.0370380866$  \\
\hline $n=6$ & 12 & $0.9644587833$  & $0.8951393273$ \\
\hline $n=8$ & 16 & $0.9815354429$ & $1.0931988113$ \\
\hline  $n=10$& 20 & $0.9728969059$  & $0.8486421716$  \\
\hline $n=12$ & 24 & $0.9757472737$  & $1.1635486772$ \\
\hline
\end{tabular}
\label{realresultsx1}
\end{table}

Finally, let us show the $L_0$ level truncation analysis of the
gauge invariant overlap for Jokel's solution. In order to expand
the string field (\ref{Soljokelg1}) in the state space of Virasoro
$L_0$ eigenstates, we need to write this string field as follows
\cite{Arroyo:2017mpd}
\begin{eqnarray}
\label{JokelL0} \hat{\Psi}_{\text{Jok}} = \int_{0}^{\infty} dt
\frac{e^{-t} r \sin ^2\left(\frac{\pi }{2 r}\right) \left(2 \pi
r-r\sin \left(\frac{\pi }{r}\right)+\pi
   \right)}{16 \pi ^2} \widetilde{U}_{r} \Big(c\big(-\frac{2 \tan \left(\frac{\pi  t}{2 r}\right)}{r}\big)+c\big(\frac{2 \tan \left(\frac{\pi  t}{2
   r}\right)}{r}\big)\Big) \nonumber \\
 +  \int_{0}^{\infty} dt \sum_{k=1}^{\infty} \frac{e^{-t}(-1)^{k+1} 2^{2 k-3} \left(\frac{1}{r}\right)^{2 k-3} \sin ^4\left(\frac{\pi }{2 r}\right)}{\pi ^2 \left(4
   k^2-1\right)} \widetilde{U}_{r}
 b_{-2k} c\big(-\frac{2 \tan \left(\frac{\pi  t}{2 r}\right)}{r}\big)c\big(\frac{2 \tan \left(\frac{\pi  t}{2
   r}\right)}{r}\big) \nonumber \\
   + \int_0^{\infty} dt_1 \int_0^{\infty} dt_2  \frac{e^{-t_1-t_2}(1+t_1+t_2)^2 \cos ^2\left(\frac{\pi  (t_2-t_1)}{2 (1+t_1+t_2)}\right)}{8 \pi }
   \widetilde{U}_{1 + t_1 + t_2} c\left(\frac{2 \tan \left(\frac{\pi  (t_2-t_1)}{2
   (1+t_1+t_2)}\right)}{1+t_1+t_2}\right), \;\;\;\;\;\;\;\;
\end{eqnarray}
where $r=1+t$.

By writing the $c$ ghost in terms of its modes $c(z)=\sum_{m}
c_m/z^{m-1}$ and employing equations (\ref{uux2}) and
(\ref{JokelL0}), the string field $\hat{\Psi}_{\text{Jok}}$ can be
readily expanded and the individual coefficients can be
numerically integrated. For instance, let us write the expansion
of $\hat{\Psi}_{\text{Jok}}$ up to level four states
\begin{align}
\label{JokelL02} \hat{\Psi}_{\text{Jok}}  = & +0.45457753 c_1
|0\rangle + 0.17214438 c_{-1} |0\rangle -0.03070678 L_{-2} c_{-1}
|0\rangle -0.01400692 b_{-2} c_0 c_1 |0\rangle \nonumber \\ &
-0.00605891 L_{-4} c_1|0\rangle + 0.02033379 L_{-2} L_{-2}
c_1|0\rangle + 0.16194599 c_{-3}|0\rangle  \nonumber \\ &
-0.00976204 b_{-2} c_{-2} c_{1}|0\rangle -0.01053192 L_{-2}
c_{-1}|0\rangle +
 0.00976204 b_{-2} c_{-1}c_{0}|0\rangle \nonumber
\\ &  + 0.00465417 b_{-4} c_{0} c_{1}|0\rangle -0.00308797 L_{-2} b_{-2} c_{0}
c_{1}|0\rangle.
\end{align}

In order to compute the gauge invariant overlap using the $L_0$
level truncation scheme, we perform the replacement
$\hat{\Psi}_{\text{Jok}} \rightarrow z^{L_0}
\hat{\Psi}_{\text{Jok}}$ and then using the resulting string field
$z^{L_0} \hat{\Psi}_{\text{Jok}}$, we define
\begin{eqnarray}
\label{gaugedependzjokel} \langle \mathcal{V} |
\hat{\Psi}_{\text{Jok}} \rangle(z) \equiv  \langle \mathcal{V} |
z^{L_0} \hat{\Psi}_{\text{Jok}} \rangle.
\end{eqnarray}
It turns out that if we naively set $z=1$ in
(\ref{gaugedependzjokel}), we obtain a non-convergent result,
therefore in the case of Jokel's solution, we are also required to
use Pad\'{e} approximants.

We have considered the string field $ \hat{\Psi}_{\text{Jok}}$
expanded up to level twenty four states, so that we obtain a
series expansion for (\ref{gaugedependzjokel}) truncated up to the
order $z^{23}$. The explicit expression for the gauge invariant
overlap, truncated up to this order, is given by
\begin{align}
\langle \mathcal{V} |  \hat{\Psi}_{\text{Jok}} \rangle(z) = \Big[
& \frac{ 0.71404871}{z} + 0.27040377 z -0.11286698 z^3
 +0.03411771 z^5 \nonumber \\
&+0.133033393978 z^7 +0.0051412823 z^9
 -0.17797842572 z^{11} \nonumber \\
&+0.00302494385 z^{13} + 0.24163840461 z^{15} +
 0.0057271144 z^{17} \nonumber \\
 \label{jokelgaugeexpanlevel11} &-0.30732930326 z^{19} -0.00086271048 z^{21} +
 0.3881263427 z^{23} \Big] \mathcal{C}_\mathcal{V}.
\end{align}
Starting from this expression (\ref{jokelgaugeexpanlevel11}), we
have computed the value of the gauge invariant overlap using
Pad\'{e} approximants of order $P^n_{1+n}(z)$. Since these
computations are similar to the ones developed in the case of
Erler-Schnabl's solution, at this point we only present the
results which are shown in table \ref{jokelrealresultsx1}. We
observed that the value of the gauge invariant overlap evaluated
using Pad\'{e} approximants confirms the expected analytic result.

\begin{table}[ht]
\caption{The Pad\'{e} approximation for the value of the gauge
invariant overlap $\langle \mathcal{V} | z^{L_0}
\hat{\Psi}_{\text{Jok}} \rangle$ divided by
$\mathcal{C}_\mathcal{V}$ and evaluated at $z=1$. The third column
shows the $P_{1+n}^{n}$ Pad\'{e} approximation. In the last
column, $P^{2n}_1$ represents a trivial approximation, a naively
summed series. At each line, we have considered the string field
expanded up to level $2n$ states.} \centering
\begin{tabular}{|c|c|c|c|}
\hline
  & Level  & $P^{n}_{1+n}$ &  $P^{2n}_1$   \\
    \hline $n=0$ & 0 & $0.7140487176$ & $0.7140487176$ \\
\hline  $n=2$& 4 & $0.9048229924$  & $0.8715855076$ \\
\hline $n=4$ & 8 & $0.9042818456$  & $1.0387366169$  \\
\hline $n=6$ & 12 & $0.9506363141$  & $0.8658994735$ \\
\hline $n=8$ & 16 & $0.9699784236$ & $1.1105628220$ \\
\hline  $n=10$& 20 & $0.9642533690$  & $0.8089606332$  \\
\hline $n=12$ & 24 & $0.9715134811$  & $1.1962242655$ \\
\hline
\end{tabular}
\label{jokelrealresultsx1}
\end{table}

\section{Summary and discussion}

Through analytical and numerical techniques, we have evaluated the
gauge invariant overlap for solutions within the $KBc$ algebra. In
order to numerically analyze the gauge invariant overlap, we have
used two types of expansions for the truncated solutions, namely,
the curly $\mathcal{L}_0$ and the Virasoro $L_0$ level expansions.

We have shown that when we expand a solution $\Psi$ in the basis
of curly $\mathcal{L}_0$ eigenstates, the resulting expression for
the gauge invariant overlap $\langle I | \mathcal{V}(i) | \Psi
\rangle$ is given in terms of a finite series and so the use of
Pad\'{e} approximants was not necessary. This is a quite generic
result provided that the solution belongs to the state space
constructed out of elements in the $KBc$ algebra. As explicit
examples, we have presented the results for the case of Schnabl's,
Erler-Schnabl's and Jokel's solutions.

Regarding the Virasoro $L_0$ level truncation analysis of the
gauge invariant overlap for Erler-Schnabl's and Jokel's solutions,
we have shown that the resulting expressions are given in terms of
divergent series which nevertheless using Pad\'{e} approximants
can be numerically evaluated. These results are in contrast to the
case of Schnabl's original solution where the expression of the
gauge invariant overlap obtained from Virasoro $L_0$ level
truncation computations becomes a convergent series, therefore, in
that case \cite{Kawano:2008ry}, there was no need for using
Pad\'{e} approximants.

Our original motivation for studying the level truncation analysis
of the gauge invariant overlap has been to prepare a numerical
background to analyze more cumbersome solutions, such as the
multibrane solutions \cite{Murata:2011ep}, however there are
problems that can arise when using the $KBc$ algebra to construct
such solutions, for instance, depending on the regularization used
to define the solutions, the analytic computation of the energy
and the gauge invariant overlap becomes ambiguous
\cite{Hata:2012cy,Masuda:2012cj}, moreover, these solutions are
not well defined when expanded in the basis of Virasoro $L_0$
eigenstates \cite{Murata:2011ex}.

With the hope of constructing well-behaved solutions other than
the tachyon vacuum, recently, the $KBc$ algebra has been extended
to a larger algebra given as a string field representation of the
Virasoro algebra \cite{Mertes:2016vos}. Since the evaluation of
the gauge invariant overlap is simpler than the energy, it should
be nice to extend the results presented in our work in order to
compute the gauge invariant overlap for solutions constructed
within the proposed Mertes-Schnabl's algebra.

Finally, we would like to comment that the evaluation of the gauge
invariant overlap can be generalized for solutions in the context
of superstring field theory
\cite{Arefeva:1989cp,Berkovits:1995ab,Berkovits:1998bt}. For
instance, we should analyze the gauge invariant overlap for
solutions constructed out of elements in the so-called
$GKBc\gamma$ algebra introduced in references
\cite{Arroyo:2010fq,AldoArroyo:2012if,Arroyo:2013pha,Arroyo:2016ajg,Erler:2013wda}.
The analytic computation of this gauge invariant quantity has
already been presented for some particular solutions
\cite{Erler:2007xt,Gorbachev:2010zz,Erler:2010pr}, however it
remains the numerical analysis.

\section*{Acknowledgements}
I would like to thank Ted Erler and Matej Kudrna for useful
discussions.


\end{document}